\documentclass[fleqn,usenatbib]{mnras}

\usepackage{newtxtext,newtxmath}

\usepackage[T1]{fontenc}
\usepackage{ae,aecompl}

\usepackage{graphicx}	
\usepackage{amsmath}	
\usepackage{amssymb}	
\usepackage{epsfig}     

\title[Reply to Comment on ``Gravitational Waves From Ultra- Short Period Exoplanets"]{Reply to Comment on ``Gravitational Waves from ultra-short period exoplanets"}

\author[Cunha, Silva \& Lima]{
J. V. Cunha,$^{1}$\thanks{E-mail: jvital.cunha@ect.ufrn.br}
F. E. Silva,$^{1}$\thanks{E-mail: edsonsilva@ect.ufrn.br}
J. A. S. Lima$^{2}$\thanks{E-mail: jas.lima@iag.usp.br}
\\
$^{1}$Escola de Ci\^encias e Tecnologia, UFRN, 59078-970, RN, Brazil\\
$^{2}$Departamento de Astronomia, Universidade de S\~ao Paulo, 05508-900 SP, Brazil}

\date{Accepted XXX. Received YYY; in original form ZZZ}
\pubyear{2017}
\begin{document}
\label{firstpage}
\pagerange{\pageref{firstpage}--\pageref{lastpage}}
\maketitle
\begin{abstract}
\vskip 0.1cm
\noindent Wong et al. (2018) recently performed an encouraging criticism to our paper ``Gravitational waves from ultra-short period exoplanets" (Cunha, Silva \& Lima 2018) exploring  the potentialities of a subset of exoplanets with extremely short periods (less than 80 min) as a possible scientific target to the planned space-based LISA observatory. Here we call attention to some subtleties and limitations underlying the basic criticism  which in our view were not properly stressed in their comment. Particularly, simple estimates show that  a  sphere  encircling the Earth with a radius of 250 pc may accommodate a population $ \sim 10^{4}$  ultra-short period exoplanets with characteristic strain of the same order or higher than the ones analyzed in our paper. This means that the question related to the gravitational wave pattern of ultra-short period exoplanets may be surpassed near future by the LISA instrument with new and more definitive data. 

\end{abstract}



\vskip 0.5cm
In a recent article (Cunha, Silva \& Lima 2018), we have advocated that the quadrupolar emission of gravitational waves (GWs) from binary systems formed by nearby exoplanets of ultra-short period and their parent stars could be detected near future by the Laser Interferometer Space Antenna (LISA) observatory. Such a conjecture was supported by the existence  of three ultra-short period exoplanets, actually, the shortest subset known to date, namely: GP Com b, V396 Hya b and J1433 b whose periods are nearly 46, 65 and 78 minutes, respectively. 

Nevertheless,  Wong et al. (2018) correctly pointed out that our conclusions were derived based on the original ``classical" LISA sensitivity curve ignoring ``galactic noise confusion" (Larson et al. 2000). Further, by adopting the latest updated curve (Cornish \& Robson 2018), they concluded that our analysis suggesting the possible detectability of such objects through GWs was, at best, inconclusive. One exception was open to the the exoplanet {\it GP Com b} whose  detection,  based on the current LISA curve, was roughly estimated to be possible only after nearly 8 years mission.

The basic reasons outlined by the authors for explaining the different conclusions of both analyses were the following:  (i) the existence of a stochastic GW background ({\it SGWB}) from binary systems, the so-called  ``galactic noise confusion" ({\it GNC}),  potentially degrading the ability of LISA for detecting this kind of  sources, and (ii) the low integration time (one year) associated to a {\it signal-to-noise ratio} ({\it SNR}) below the threshold which for nearly monochromatic sources like  binary systems is usually taken to be $SNR \geq 5$ (see their Table I).  

To begin with, let us observe that the present estimates of the {\it GCN} from binary systems suffers from several drawbacks related to astrophysical uncertainties, such as the star formation rate and the form of its initial mass function and the same is valid for ultra-short period exoplanets. Actually, in the absence of a detectable sign there are also uncertainties about  how the confusion limit is clearly defined from an observational viewpoint. More important perhaps, the galactic confusion is not stationary. It diminishes rapidly in the course of time,  as long as the mission progresses and more foreground sources can surely be removed (Cornish 2004). Hence, any estimate of the {\it SGWB} at present should at best be considered as an educated guess because the abundance of binary  galactic  GW emitters is indeed unknown.  We are not claiming here that the {\it GNC} should be disregarded in the ongoing and future analyses. However, we fully agree that the question related to the {\it SNR}  and the related observation time is the most serious  difficulty to be overcome. 

 \begin{table*}
\centering
\caption{Gravitational wave parameters of three exoplanetary systems of extremely short periods  as discussed by Cunha, Silva \& Lima (2018) and Wong {\it et al}. (2018). In the last line we see the corresponding values for a hypothetic member (exoplanet {\bf X b}) of the ultra-short period exoplanetary population (see text).} 
\begin{tabular} {lcccccccccc} \hline
Planet & $m_1$ & $m_2$ &  P & {\it a} & $D_{L}$ & $f_{GW}$ & $h_c$ & SNR & SNR \\ 
 & $({M_{\odot}})$ & $(M_J)$ & (min) & (au) & $(pc)$ & $(10^{-4} Hz)$ & $ (10^{-20})$ & ($2$ yr) & ($4$ yr)
\\ \hline
GP Com b  & $0.435$	& $26.2$ & $46.6$ & $0.0015$ & $75$  & $7.2$ & $3.44$ & $2.12$ & $3.47$ \\
V396 Hya b & $0.345$ &	$18.3$ &  $65.1$ & $0.0017$ & $77$ & $5.1$ & $1.35$ & $0.60$ & $0.91$ \\ 
J1433 b & $0.8$ &	$57.1$ &  $78.1$ & $0.0027$ & $226$  & $4.3$ & $2.03$ & $0.75$ & $1.11$ \\\hline
{\bf X b} & $\mathbf{0.5}$ & $\mathbf{40.0}$ &  $\mathbf{30.0}$ & $\mathbf{0.0012}$ & $\mathbf{50}$ & $\mathbf{11.1}$ & $\mathbf{14.2}$ & $\mathbf{12.50}$& $\mathbf{25.95}$ \\\hline
\end{tabular}
\end{table*}

Now, let us discuss the {\it SNR} for ultra-short period exoplanets  and its possible detection by LISA. In our viewpoint the challenging question is to justify (beyond reasonable doubt) whether  ultra-short period exoplanets are interesting scientific targets for LISA. 

In Table I we present a short list of systems containing  ultra-short period exoplanets including their masses and orbital parameters, planet period (P),  major semi axis ({\it a}) distance to the Earth ($D_L$), GW frequency ($f_{GW}$), the calculated characteristic strain $h_{c}$ and the associated {\it SNR} for 2 and 4 years of observation time. 

The first three lines in Table I  stand for the super Jupiters already observed and discussed in the paper and comment.  Apart the  {\it SNR} of J1433 b for 2 years in the comment, our results in the last two columns agree with the ones presented by Wong et al. (2018). Note also the presence of a hypothetical exoplanet  {\bf X\,b} (fourth line), assumed to be  a regular super Jupiter because its orbital parameters suggest that it is  not a special or  extreme object. It is easy to show that such a possible candidate of the class has {\it SNR = 12.50} and {\it SNR = 25.95} for observation times of 2 and 4 years, respectively. All the {\it SNR} results were averaged over sky/polarization and inclination. Naturally, less massive planets than {\bf X\,b} with shorter periods and distances  are not forbidden, and different {\it SNR} would be obtained for reasonable choices of the basic parameters.  In particular, like verification and other galactic binaries (Cornish \& Robson 2018), such possibilities reinforce the idea that a cloud of ultra-short period exoplanets may occupy the left upper corner of the current sensitivity LISA curve around a frequency of $10^{-3}Hz$ (see Figure 1 and discussion next).  

In principle,  objects  with similar parameters  would play a special role as a scientific target for LISA mission.  Therefore, at this point it is natural to ask about their abundance. In other words:  {\it Is there a class of such objects at the Earth neighbourhood?} In our view this is very probable and the basic reasons are justified below.

The average density of stars in the solar neighborhood is about $0.14$\, stars.$pc^{-3}$. Hence, within a spherical shell encircling the Earth with a radius of 250 pc (nearly 25 pc higher than the distance to the farthest ultra-short period exoplanet known, J1433 b in Table I), one may expect at least the existence of $10^{7}$ stars. Note that the distance also need to be small in order to present a reasonable characteristic strain. On the other hand, at present, nearly 3000 stars with exoplanets from which only three of them own ultra-short period exoplanets already identified.  Such a result can be translated to a naive but conservative estimate of nearly $10^{4}$ objects of this class up to 250 pc (we are neglecting the remaining 11 candidates from  Table I in Cunha, Lima \& Silva 2018).   
  
In Figure 1, we display the classical (red line) and current (black line) LISA sensitivity curves where the latter already includes the {\it GCN}. Basically, they are the same curves presented by Wong et al. (2018) but now including the hypothetical  {\bf X\,b} exoplanet  with   {\it SNR=12.50} for two years observation time (see also Table I). 

In light of the above considerations, it is very temptative to conjecture the existence of a class of super Jupiters filling the left upper corner of the space parameter as indicated by the grey elipsoidal region in Figure 1. The possible existence of this  class was explicitly suggested  in our work. As it appears, this class is only partially contaminated by the {\it GCN}, and, as such, a large portion  of it is potentially detectable by LISA. This central point underlying our work in our view was not properly stressed in previous papers. 

\begin{figure}
\centerline{\epsfig{figure=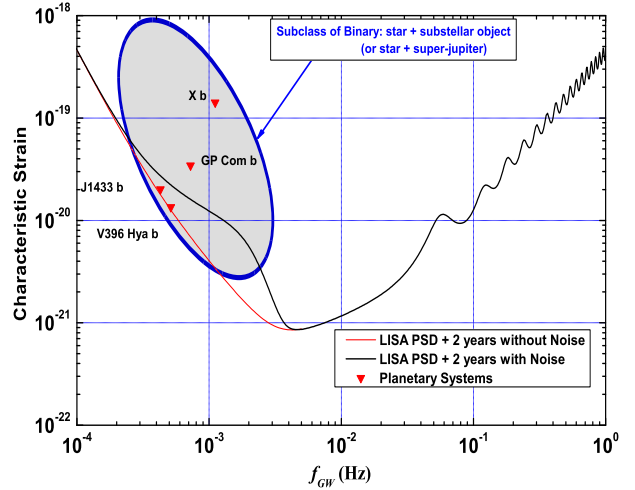,width=3.2truein,height=2.6truein}}
\caption{$h_c$ versus frequency for exoplanets listed on Table 1. Black and red lines are, respectively,  the current and the classical LISA  sensitivity curves with and without ``galactic confusion noise" for 2 years of integration time. The three already detected exoplanets (plus the hypothetical {\bf X b}) suggest a conservative population of nearly 10,000 objects encircling the Earth up to 250 pc (see text), a  portion of them detectable by LISA.}
\end{figure}
 
 Summarizing, after nearly 5,000 thousand planets already identified, someone may consider that the golden age of hunting planets through optical methods (transiting time, radial velocity, etc.)  is coming to an end. However, the above numbers suggest that such a frontier may be replaced by an equally challenging hunting process now based on gravitational waves searching for nearby ultra-short period exoplanets.  Finally, we also stress that the present discussion involving galactic noise and the possible detection of ultra-short period exoplanets through LISA observatory may be surpassed by new and definitive data in the near future.  
\section*{Acknowledgements} We thank Jos\'e Carlos Neves de Ara\'ujo and Carlos Eduardo Pellicer for useful discussions. JASL is grateful to CAPES (PROCAD 2013), FAPESP (LLAMA project) and a fellowship from CNPq.

\bsp
\label{lastpage}
\end{document}